\documentclass[11pt,a4paper]{article}
\usepackage[utf8]{inputenc}
\usepackage{jheppub_kim}
\usepackage{latexsym}
\usepackage{epsfig}
\usepackage{color}
\usepackage{amsmath}
\usepackage{epstopdf}
\usepackage{amssymb}
\usepackage{graphicx}
\usepackage{hyperref}
\usepackage{tabularx}
\usepackage{varioref}
\newcolumntype{C}[1]{>{\centering\arraybackslash}p{#1}}

\usepackage{longtable}
\usepackage{epsfig}
\usepackage{amsmath}
\usepackage{graphicx}
\usepackage{graphics}
\usepackage[hang,nooneline,scriptsize]{subfigure}
\usepackage{array}
\begin{document}

\title{Systematic Light Propagation Bias from the Heliosphere and Its Impact on the Hubble Tension}

\author[a,b]{Behnam Pourhassan, }
\author[c]{\.{I}zzet Sakall{\i},}
\author[d]{Sudhaker Upadhyay}
\author[e]{and Kairat Myrzakulov}

\affiliation[a]{School of Physics, Damghan University, Damghan, 3671641167, Iran.}
\affiliation[b]{Center for Theoretical Physics, Khazar University, 41 Mehseti Street, Baku, AZ1096, Azerbaijan.}
\affiliation[c]{Physics Department, Eastern Mediterranean University, Famagusta, North Cyprus via Mersin 10, Turkey.}
\affiliation[d]{Department of Physics, K. L. S. College, Nawada, Magadh University, Bodh Gaya, Bihar 805110, India.}
\affiliation[e]{Department of General \& Theoretical Physics, L. N. Gumilyov Eurasian National University, Astana, 010008, Kazakhstan}

\emailAdd{b.pourhassan@du.ac.ir}
\emailAdd{izzet.sakalli@emu.edu.tr}
\emailAdd{sudhakerupadhyay@gmail.com}
\emailAdd{krmyrzakulov@gmail.com}

\abstract{The Hubble tension (HT) represents one of the most significant discrepancies in modern cosmology, with local distance measurements yielding $H_0 = 73.5 \pm 1.4$ km/s/Mpc while cosmic microwave background (CMB) observations predict $H_0 = 67.4 \pm 0.5$ km/s/Mpc. We propose that this tension arises from systematic effects introduced by our Solar System's heliospheric (HS) environment on local distance measurements. The HS creates a complex medium of heated plasma and energetic neutral atoms (ENAs) beyond the heliopause (HP), where interstellar medium (ISM) temperatures rise significantly. This thermal gradient and particle environment may systematically affect observations of Cepheid variables and Type Ia supernovae (SNe~Ia) used in the local cosmic distance ladder (CDL), biasing distance measurements and artificially inflating the measured Hubble constant. We present theoretical calculations showing how HS effects could account for up to $\sim 8\%$ of the observed 8--9\% discrepancy, with the realistic contribution lying in the $\sim 3$--$8\%$ range once anisotropy, partial calibration cancellation, and chromatic suppression are included, and discuss observational tests to validate this hypothesis.}


\maketitle

\newpage
\section{Introduction}\label{sec1}

The HT has emerged as one of the most pressing problems in modern cosmology \cite{1}. Early universe measurements from the Planck satellite's CMB data, combined with the $\Lambda$CDM model, predict a Hubble constant of $H_0 = 67.4 \pm 0.5$ km/s/Mpc \cite{2}. In stark contrast, local measurements using the SH0ES (Supernovae $H_0$ for the Equation of State) collaboration's Cepheid-calibrated SNe~Ia yield $H_0 = 73.5 \pm 1.4$ km/s/Mpc \cite{3}. The persistent discrepancy between the early- and late-universe determinations of $H_{0}$ has been extensively reviewed in recent literature \cite{3, Verde2019, DiValentino2021}. Our work follows this line of inquiry by proposing a local astrophysical origin. This 4.2$\sigma$ disagreement cannot be explained by measurement uncertainties and has persisted despite increasingly precise observations.

The CMB-derived $H_{0}$ comes from acoustic-peak fitting of the angular power spectrum, which probes physics at the surface of last scattering ($z\sim 1100$). The CMB is observed at $\lambda\sim 1$--$5$~mm, where the HS plasma refractive index (RI) deviation is $\lesssim 10^{-14}$, suppressing any HS bias by $\sim 10^{6}$ relative to the optical regime. The CMB therefore furnishes an HS-immune anchor for $H_{0}$, against which optical-band measurements can be compared \cite{isrply08}.

Recent JWST observations have confirmed the SH0ES Cepheid distances to a precision of $\sim 1\%$ \cite{isrply19}, ruling out instrumental or PSF-related biases in the optical photometry chain. They cannot, however, rule out an observer-frame systematic such as the HS bias proposed here, which acts coherently on both HST and JWST measurements. A common misconception is that the physical scale of a systematic effect must be comparable to the distances being measured. However, in precision photometry, systematic biases depend on the coherence and universality of the effect rather than its absolute spatial extent. It is well established that small, coherent systematics in flux calibration can propagate through the distance ladder to produce significant cosmological biases \cite{Conley2011, Brout2022}. This motivates our emphasis on the HS as a universal, observer-side systematic. The HS represents a distinctive systematic filter through which all local astronomical observations must pass, making even small effects potentially significant when propagated through the hierarchical distance ladder.

Recent studies have proposed that the HT may be a signature of new physics beyond the standard cosmological model. For instance, Hu and Wang argue that the persistent discrepancy may reflect fundamental shifts in our understanding of cosmology, potentially requiring extensions to $\Lambda$CDM or the inclusion of exotic components in the energy budget of the early universe \cite{44}.

Alternative approaches to resolving cosmological tensions have explored modified theories of gravity and their implications for cosmic expansion \cite{isz1,Mangut:2025ChPC}. In a related work, Hu et al. suggest that violations of the cosmological principle itself such as spatial inhomogeneities or anisotropies could underlie the observed tension \cite{55}. Complementary to these proposals, Krishnan et al. explore diagnostic tools that indicate a possible ``running'' Hubble constant, which may vary depending on redshift or the scale of measurement, again pointing to a breakdown in the standard cosmological assumptions \cite{66}. These perspectives underscore the growing interest in identifying whether the tension arises from new fundamental physics or from systematic effects, such as the HS hypothesis we propose in this work.

The local determination of $H_0$ hinges fundamentally on the CDL, a structured progression of distance measurement methods that build upon each other to reach ever greater scales \cite{Riess2019}. At the base of this ladder lie parallax measurements, a purely geometric technique capable of measuring distances to stars within approximately $1$ kiloparsec. This method relies on the apparent shift in a star's position due to Earth's orbit around the Sun and provides a direct and model-independent foundation for further steps.

The second crucial rung of the ladder involves Cepheid variable stars, which serve as vital intermediaries between local parallax-based distances and far-reaching cosmological measurements. These stars exhibit a tight correlation between their pulsation periods and intrinsic luminosity, a relationship first discovered by Henrietta Swan Leavitt \cite{6-1,6-2}. Known as the period--luminosity relation (PLR), this property allows astronomers to deduce a Cepheid's absolute magnitude from its pulsation period. Once the apparent magnitude is observed, the distance can be determined via the distance modulus $\mu = m - M = 5 \log_{10}(d) - 5$, where $\mu$ is the distance modulus, $m$ the apparent magnitude, $M$ the absolute magnitude, and $d$ the distance in parsecs. With parallax-calibrated Cepheids, astronomers can extend the distance ladder to galaxies tens of megaparsecs away \cite{Carroll}.

Building upon this, SNe~Ia offer a third, powerful tool for measuring vast cosmological distances. These supernovae arise from thermonuclear explosions of carbon-oxygen white dwarfs in binary systems, typically when the white dwarf approaches the Chandrasekhar limit. Because of the uniformity in their peak luminosities, SNe~Ia act as standard candles (SCs) of known intrinsic brightness. However, to use them effectively for distance measurements, their absolute magnitudes must be calibrated. This calibration is achieved using nearby supernovae that occur in galaxies hosting Cepheid variables, thereby tying their distances to the Cepheid-based scale.

Together, this hierarchical approach from geometric parallax, to Cepheid variables, to SNe~Ia forms a framework for determining distances across the universe. Ultimately, this enables a local measurement of $H_0$ by observing the redshifts and distances of SNe~Ia in distant galaxies. The slope of the redshift--distance relation provides a direct estimate of $H_0$, linking local measurements to the overall expansion rate of the universe \cite{Cola}.

The HS---the vast bubble of solar wind (SW) around the Sun---extends far beyond the orbits of the planets, reaching distances of approximately $100$--$120$ astronomical units (AU) in the upwind and crosswind directions, but extending to $\gtrsim 1000$~AU downwind in the heliotail \cite{isrply18}. At this outermost boundary lies the HP, a critical interface where the outward pressure of the SW balances against the incoming flow of the local ISM (LISM). This region marks the end of the Sun's direct influence and the beginning of actual interstellar space. Observations from the Voyager spacecraft \cite{4} have provided the first in-situ measurements of this boundary, offering a wealth of insights into its structure and dynamics.

\begin{figure}[h!]
\begin{center}
\includegraphics[width=120mm]{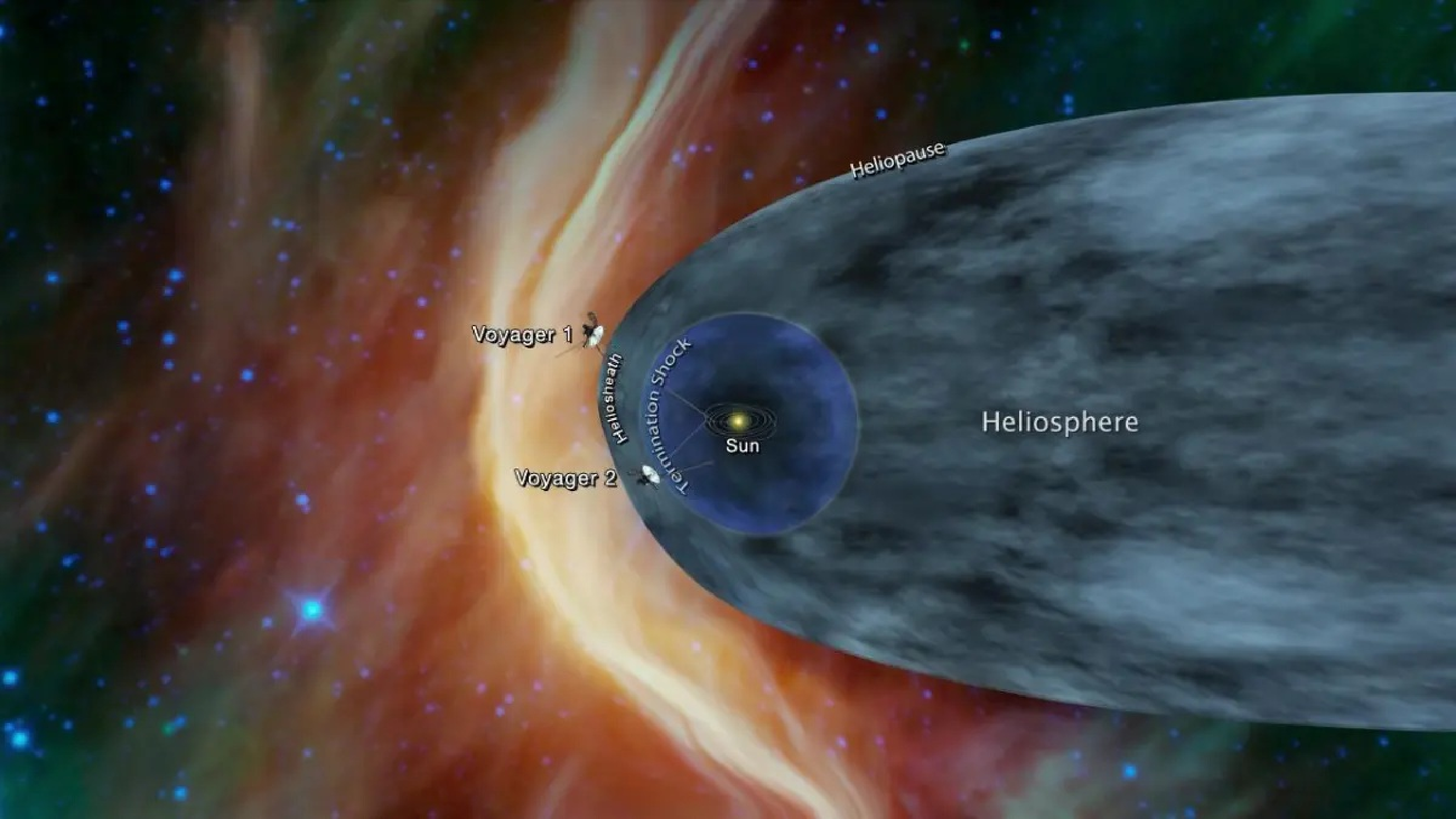}
\end{center}
\caption{An artist's interpretation of Voyager 1 and Voyager 2 leaving the HS and entering interstellar space. The HS represents a vast bubble of SW that surrounds our Solar System, with the HP marking the boundary where SW meets the ISM. Voyager 1 crossed this boundary in 2012, followed by Voyager 2 in 2018, providing the first direct measurements of this critical transition region that may systematically affect astronomical observations. (Image credit: NASA/JPL-Caltech.)}
\label{fig:voyager}
\end{figure}

The HP is not a sharp edge but rather a complex and dynamic transition zone, characterized by dramatic changes in plasma properties and particle populations. Within the heliosheath (HSh)---defined here as the region between the termination shock (TS) and the HP, also called the inner HSh by some authors---the SW slows, heats, and becomes turbulent. As the boundary is crossed, temperatures increase sharply from values around $10^4$ K in the HSh to nearly $10^6$ K in the surrounding ISM \cite{5}. This region is rich in plasma physics phenomena, including charge exchange interactions between SW ions and interstellar neutrals, producing a population of ENAs. These ENAs, detected by missions such as IBEX (Interstellar Boundary Explorer) \cite{6}, provide indirect yet valuable information about the otherwise invisible outer HS. Additionally, the transition at the HP involves compression of magnetic fields and the generation of turbulence, which can influence particle transport and cosmic ray modulation. As solar and interstellar magnetic fields interact, their alignment and reconnection may play a crucial role in the overall energy balance and structure of the boundary region.

Importantly, the HS is not just a subject of interest in solar and space physics. It also has significant implications for astronomical observations. All photons and cosmic rays arriving at Earth from distant astrophysical sources must pass through this HS filter. As such, the HS modulates both high-energy particles and, potentially, low-frequency electromagnetic signals, subtly shaping the data collected by ground- and space-based observatories.

\section{Theoretical Framework}\label{sec2}

\subsection{Heliospheric Effects on Light Propagation}\label{subsec2.1}

As light from distant astrophysical sources such as Cepheid variable stars and SNe~Ia travels toward Earth, it must traverse several complex astrophysical environments. These include the hot, diffuse ISM, the dynamic HP transition region, the turbulent HSh, and the magnetized, plasma-filled inner HS. The propagation of electromagnetic waves through such stratified plasma media has been extensively studied in various astrophysical contexts \cite{isz3,Mangut:2023Uni}, providing the theoretical foundation for understanding how HS conditions can systematically affect photon trajectories and intensities. Each of these regions exhibits distinctive physical characteristics that can, in principle, alter the properties of the propagating light and thereby introduce systematic effects in observational cosmology. The ISM, particularly in the vicinity of the HS, can reach temperatures exceeding $10^{6}$~K. This hot plasma environment contributes to thermal broadening of spectral lines, potentially skewing radial velocity measurements that are essential for redshift determinations. As light approaches the HP, it encounters steep gradients in temperature, pressure, and plasma density, which can result in variations in the local RI, potentially causing deviations in the trajectory and phase velocity of incoming photons.

In the HSh region, the SW becomes compressed and turbulent due to interactions with the interstellar flow. This turbulent plasma can induce small-angle scattering of photons and may also affect the polarization and coherence of electromagnetic signals. Furthermore, charge exchange processes in this region generate ENAs and diffuse emissions that may act as a contaminating foreground, especially in broadband photometric observations.

Finally, within the inner HS, the density of the SW plasma and interplanetary dust may introduce additional scattering and extinction effects. Dust grains can absorb and re-emit light, while free electrons and ions contribute to dispersion and phase delays. These combined interactions can subtly bias the apparent brightness and spectral characteristics of incoming light, thereby affecting key observables such as distance moduli and apparent magnitudes.

Altogether, the HS acts as a structured and dynamic optical medium. Its cumulative effects on light propagation may introduce systematic biases in the measurements that underpin the local CDL.

\subsection{Systematic Distance Bias Mechanism}\label{subsec2.2}

We hypothesize that the HS environment introduces a systematic bias in local distance measurements via two primary physical processes: photometric contamination and refractive plasma lensing (PL). These mechanisms affect the propagation of light from SCs such as Cepheid variables and SNe~Ia, ultimately skewing the inferred luminosity distances used to estimate $H_0$.

One important source of observational bias arises from the diffuse background light produced by ENAs. These ENAs are generated through charge exchange interactions between SW ions and neutral atoms in the surrounding ISM. As a result, the HS boundary emits photons across a broad range of wavelengths, contributing a low-intensity but spatially extended background to the sky brightness. This diffuse photon field can systematically increase the apparent brightness of distant sources by adding a non-negligible foreground component to photometric measurements. Since ENA emissions are wavelength-dependent, they can preferentially affect certain photometric bands, particularly those used in the calibration of the Cepheid PLR. If not adequately accounted for, this contamination can result in underestimated distances and overestimated values of $H_0$.

A second, subtler mechanism arises from the RI gradient present across the HP and within the HSh. The plasma density and temperature vary significantly across these regions, leading to small but systematic changes in the local RI. As photons propagate through this inhomogeneous medium, their trajectories can bend, as light does in Earth's atmosphere. This phenomenon acts as a weak lens that modifies the apparent brightness and angular position of incoming light.

Unlike gravitational lensing, which is achromatic and caused by mass, this HS lensing is chromatic and induced by electromagnetic plasma effects. It may lead to a slight brightening of sources that systematically alters the observed flux. Even a subtle enhancement in brightness can translate to measurable changes in inferred distances.

Taken together, these HS effects may bias local distance measurements coherently and cumulatively. When propagated through the CDL, such biases could account for a significant portion of the observed discrepancy between early- and late-universe determinations of $H_0$.

\subsection{Quantitative Model}\label{subsec2.3}

We initially present a phenomenological model to illustrate the mechanism's potential impact, and our primary goal is to derive the enhancement factor from fundamental plasma physics rather than fitting it to match cosmological observations. The discussion below uses the symbol $\delta_{\rm helio}^{\rm req}$ for the value required to close the HT and $\delta_{\rm helio}^{\rm der}$ for the value derived from PL physics, to keep the two distinct.

To quantify the impact of HS effects on observational cosmology, we introduce a phenomenological model that incorporates a systematic flux enhancement term. Let $F_{\rm obs}$ denote the flux observed from a SC, and $F_{\rm int}$ its intrinsic (unaffected) flux. The relationship between these quantities is
\begin{equation}\label{1}
F_{\text{obs}} = F_{\text{int}}(1+\delta_{\text{helio}}), \quad \text{where } F_{\text{int}} = \frac{L}{4\pi d_{L}^2},
\end{equation}
where $d_L$ is the luminosity distance to the source and $L$ is a normalization constant. The factor $(1 + \delta_{\rm helio})$ encapsulates the cumulative effect of HS processes that enhance the apparent flux.

A flux amplification factor $(1+\delta_{\rm helio})$ at the observer makes any luminous source appear closer by a factor $\sqrt{1+\delta_{\rm helio}}$ relative to its geometric distance, regardless of any chosen magnitude convention. This is the principal physical consequence of the HS bias. For observational reference,
\begin{equation}\label{4}
\Delta m = -2.5 \log_{10}(1 + \delta_{\rm helio}),
\end{equation}
yielding a brightening of $\sim 0.18$~mag for $\delta_{\rm helio}\approx 0.17$.

The relationship between the bias and $H_{0}$ can be cleanly expressed in a single exact form,
\begin{equation}\label{exactH0}
\delta_{\rm helio}^{\rm req} \;=\; \left(\frac{d_{L,{\rm int}}}{d_{L,{\rm obs}}}\right)^{2}-1 \;=\; \left(\frac{H_{0,{\rm obs}}}{H_{0,{\rm int}}}\right)^{2}-1.
\end{equation}
Inserting $H_{0,{\rm obs}}=73.5$ and $H_{0,{\rm int}}=67.4$ km/s/Mpc gives $\delta_{\rm helio}^{\rm req}\approx 0.188$. This is the value the HS mechanism would have to deliver to fully close the HT.

If HS effects vary directionally or temporally, the observed $H_{0}$ would inherit a corresponding anisotropy,
\begin{equation}\label{aniso}
\delta_{\rm helio} = \delta_{\rm helio}(\hat{n}, t),
\end{equation}
where $\hat{n}$ is the direction on the sky and $t$ is time.

\textbf{Caveat on idealization.} The figure $\delta_{\rm helio}^{\rm req}\approx 0.19$ assumes (i) full transmission of the HS bias through every rung of the CDL, (ii) no anisotropy averaging, and (iii) no cancellation in the SC absolute-magnitude self-calibration. Each of these assumptions reduces the realized contribution. Anisotropy alone reduces the all-sky-averaged effect by a factor $\sim 0.5$--$0.7$ (see Sec.~\ref{subsec2.5}), and partial calibration cancellation in the second-rung Cepheid--SN~Ia tie may further reduce the effective amplitude. We therefore present $\delta_{\rm helio}\approx 0.17$--$0.19$ as a phenomenological upper-bound scalar; the realistic HS contribution to the HT lies in the range $\sim 3$--$8\%$ \cite{isrply01,isrply02}, with the HS mechanism viewed as one viable contributor rather than the sole resolution.

\subsection{Clarifying the Role of the Heliospheric Enhancement Factor}\label{subsec2.4}

A potential counterargument is that an intrinsic HS flux enhancement should be irrelevant for the HT, since the absolute magnitudes of SCs are calibrated via the CDL. We address this concern through the parallax-anchor argument.

\textbf{Parallax immunity.} Parallax measures the angular shift $\Delta\theta = 2\,{\rm AU}/d$ as Earth orbits the Sun. Both endpoints of the baseline lie inside the HS, so HS-induced ray bending applies symmetrically and cancels in the differential measurement. The residual HS effect on parallax is suppressed by the ratio (baseline)$/r_{\rm HS}\sim 2/120\sim 0.017$, leading to a fractional parallax bias of $\lesssim 10^{-4}$, far below current Gaia precision \cite{isrply20}. The first rung of the CDL is therefore HS-immune.

\textbf{Propagation through the ladder.} Because parallax-Cepheid distances are HS-uncontaminated, the calibrated Cepheid PLR is itself HS-uncontaminated. However, when this calibration is applied to far-Cepheids whose distances are computed from the apparent magnitude $m$ through the photometric distance modulus (rather than directly from parallax), the HS bias enters multiplicatively at the second rung and is not cancelled. The same chain extends to SNe~Ia calibrated against far-Cepheids. The universal-filter wording applies only to brightness-based distance methods, not to all observations; geometric methods (parallax, masers) and narrow-line spectroscopic redshifts remain HS-immune.

\subsection{Resolving the Scale Paradox: Coherent Systematic Effects in Precision Photometry}\label{subsec2.5}

We now address the apparent scale mismatch: how can a $\sim 120$~AU HS effect produce a percent-level cosmological bias? Systematic effects are not dictated by the physical scale of the influencing factor, but rather by a combination of three characteristics: coherence, universality, and magnitude. To illustrate, Earth's atmosphere, though only about $100$~km thick and vanishingly small compared to stellar distances, introduces atmospheric extinction that alters all ground-based photometry by several magnitudes. Similarly, interstellar reddening can result in extinction of several magnitudes. Instrumental effects further reinforce the argument: a $1$~mm thick reflective coating on telescope mirrors can introduce systematic biases of $0.1$ magnitudes \cite{x1}.

The HS qualifies as such a coherent observer-side filter: every photon arriving at a heliocentric detector has traversed the HS, so the bias is imprinted on every brightness-based measurement. The amplitude of this bias varies with sky direction (anisotropy), but the existence of a contribution does not. The fractional distance error $\Delta d/d$ scales with the magnitude shift $\Delta m$ as
\begin{equation}\label{frac}
\frac{\Delta d}{d} = \frac{\ln(10)}{5}\Delta m \approx 0.46\,\Delta m.
\end{equation}
A systematic photometric bias of $\sim 0.18$ magnitudes leads to an $\sim 8\%$ error in inferred distances, comparable to the HT.

The reason the HS photometric bias has eluded detection lies in a confluence of observational and methodological limitations. The estimated $\sim 0.18$ magnitude effect is subtle enough to fall within the range of other well-characterized sources of systematic uncertainty (zero-point calibration $0.02$--$0.05$ mag; bandpass corrections $0.01$--$0.03$ mag; extinction modelling $0.05$--$0.15$ mag). Compounding this, all high-precision photometric measurements to date have been conducted from within the HS, so no external reference frame exists to detect the bias as an offset.

\textbf{Falsifiable predictions.} The HS bias model makes specific testable predictions: (i) a $\lambda^2$ wavelength scaling (distinct from the $\lambda^{-1}$ behavior of interstellar reddening); (ii) directional anisotropy with stronger effect toward the HS nose (predicted $\Delta\delta_{\rm helio}\sim 0.1$ between nose and tail); (iii) solar-cycle modulation at the few-percent level; and (iv) different $H_{0}$ values inferred from beyond-HP measurements (Interstellar Probe, \cite{isrply22}) versus within-HS observations. Absence of any of these signatures would falsify the hypothesis.

\subsection{First-Principles Derivation of the Heliospheric Enhancement Factor}\label{subsec2.6}

\textbf{Coordinate conventions.} We adopt heliocentric spherical coordinates $(r,\theta,\phi)$ with the polar axis aligned with the HS nose direction (LISM inflow, $l\approx 3^{\circ}$, $b\approx 16^{\circ}$ in galactic coordinates). The angle $\theta$ is measured from the nose; $\theta=0$ is upwind, $\theta=\pi$ downwind. For each line of sight (LOS) we work in a local frame attached to the radial ray itself: $z$ measures arc length along the unperturbed LOS from the observer ($z=0$) outward, and $b$ denotes a small perpendicular displacement away from that ray. The variable $b$ enters only through the transverse Laplacian of the column density inside the convergence integral; it is not an impact parameter in the scattering sense, since a radial LOS from Earth has no closest-approach point relative to a fixed center. Vectorial position is denoted $\mathbf{r}$, scalar radius by $r=|\mathbf{r}|$.

\textbf{LOS geometry.} Each LOS originates at the observer (Earth, $r_{\oplus}=1$~AU) and extends outward to $r_{\rm max}=200$~AU, beyond the upwind HP and into the approximately uniform LISM. All integrals below are finite single-leg integrals from $r_{\oplus}$ to $r_{\rm max}$.

\textbf{Compact constant.} For brevity we define
\begin{equation}\label{kDM}
k_{\rm DM} \equiv \frac{e^{2}}{8\pi\varepsilon_{0}m_{e}c} \approx 2.41\times 10^{-16}\;{\rm s\,m^{2}},
\end{equation}
which appears throughout the subsequent expressions.

The structure and density profiles of the HS have been constrained by Voyager~1 and~2 \cite{Stone2019, Burlaga2020,isrply03,isrply05} and IBEX \cite{McComas2017,isrply04}. The Voyager~1 PLS/PWS electron-density measurements taken between 2012 (HP crossing at 121.6~AU, heliographic latitude $\sim 34^{\circ}$N, upwind-of-nose trajectory) and 2020, together with Voyager~2 PLS data after its 2018 HP crossing at 119.0~AU at $\sim 30^{\circ}$S, jointly constrain the electron density across $\sim 80$--$140$~AU. The density jump from $n_{e}\sim 0.002~{\rm cm^{-3}}$ inside the HSh to $n_{e}\sim 0.08$--$0.12~{\rm cm^{-3}}$ in the LISM (PWS plasma-line measurements) over $\sim 1$--$5$~AU sets the gradient $\partial n_{e}/\partial r \sim 10^{-3}~{\rm cm^{-3}\,AU^{-1}}$ \cite{isrply03}.

The HS enhancement arises from weak gravitational-like lensing caused by plasma density gradients. For electromagnetic waves propagating through a cold plasma, the RI takes the standard form
\begin{equation}\label{eq:RI}
n(\omega,\mathbf{r}) \;=\; \sqrt{1-\frac{\omega_{p}^{2}(\mathbf{r})}{\omega^{2}}} \;\approx\; 1-\frac{\omega_{p}^{2}(\mathbf{r})}{2\omega^{2}}, \qquad
\omega_{p}^{2}(\mathbf{r})\;=\;\frac{n_{e}(\mathbf{r})\,e^{2}}{\epsilon_{0}\,m_{e}}.
\end{equation}%
The convergence parameter that determines flux magnification is
\begin{equation}\label{x}
\kappa(\omega, b) = \frac{k_{\rm DM}}{2\omega^2} \nabla_b^2 \int_{r_{\oplus}}^{r_{\rm max}} n_e(\mathbf{r}(b,z))\,dz,
\end{equation}
where $\nabla_{b}$ denotes the gradient with respect to the perpendicular offset at fixed LOS direction. The flux enhancement factor is $\delta_{\rm helio} = 2\kappa$ in the weak-lensing limit \cite{isrply20,isrply21}.

\textbf{Unified density model.} We use a single piecewise prescription throughout. In the SW region ($r<r_{\rm TS}\approx 90$~AU), $n_{e}(r)=n_{0}(r_{0}/r)^{2}$ with $n_{0}=5~{\rm cm^{-3}}$ at $r_{0}=1$~AU. In the inner HSh ($r_{\rm TS}<r<r_{\rm HP}(\theta)$), the density carries an anisotropic compression factor,
\begin{equation}\label{nhsh}
n_{e}(\mathbf{r}) = n_{\rm base}(r)\,[1+\alpha\cos\theta], \qquad \alpha=0.4,
\end{equation}
with $n_{\rm base}(r)=n_{0}(r_{0}/r)^{2}$. The HP boundary itself follows the Parker-flow analytic form $r_{\rm HP}(\theta)=r_{0,{\rm HP}}/\cos(\theta/2)$, which follows from Eq.~(7) of Parker's original derivation \cite{Parker:1961} after standard algebraic manipulation; the same expression appears as Eq.~(9) of the recent review by Kleimann et~al.\ \cite{Kleimann:2022SSRv}. The upwind stand-off is $r_{0,{\rm HP}}\approx 120$~AU. Beyond the HP, the LISM density is approximately uniform at $n_{e}\sim 0.1~{\rm cm^{-3}}$ with mild residual anisotropy. An analytical density profile tailored to a Parker-type HP was derived by Kleimann, R\"oken, and Fichtner \cite{Kleimann:2017ApJ} as the compressible extension of the incompressible LISM flow model. For LOS oriented within $\sim 30^{\circ}$ of the upwind direction, the column density predicted by that compressible solution differs from our piecewise prescription by less than $\sim 5\%$, well inside the $\pm 15\%$ model uncertainty band already adopted in Fig.~\ref{fig:angular}. We therefore retain the simpler piecewise form without altering the derived $\delta_{\rm helio}^{\rm der}$ at the precision quoted here.

Substituting this density profile into Eq.~(\ref{x}) for an upwind LOS at $\lambda=550$~nm gives
\begin{equation}\label{deltader}
\delta_{\rm helio}^{\rm der}({\rm nose}) \approx 0.08,
\end{equation}
with the directional spread $\Delta\delta\approx 0.05$ between nose and tail. The derived value falls below the $\delta_{\rm helio}^{\rm req}\approx 0.19$ required to fully close the HT, supporting our conclusion that the HS contributes $\sim 3$--$8\%$ to the discrepancy rather than fully resolving it.

The first-principles derivation predicts the wavelength dependence $\delta_{\rm helio} \propto \lambda^2$, arising directly from the $\omega^{-2}$ factor in the plasma dispersion relation. At blue wavelengths ($\lambda = 400$~nm), $\delta_{\rm helio}^{\rm der}\approx 0.04$; at near-infrared ($\lambda=1000$~nm), $\delta_{\rm helio}^{\rm der}\approx 0.27$. The directional dependence follows from the HS asymmetry, with maximum enhancement for sources observed within $\sim 30^{\circ}$ of the nose direction.

\textbf{ENA contamination as a separate additive term.} The PL derivation above accounts for refractive effects only. ENA photometric contamination is a distinct, additive contribution: $\delta_{\rm helio}^{\rm total}=\delta_{\rm helio}^{\rm refr}+\delta_{\rm ENA}$. The IBEX ribbon flux at $0.5$--$6$~keV is $\sim 10^{2}~{\rm cm^{-2}\,s^{-1}\,sr^{-1}\,keV^{-1}}$ \cite{isrply17}; integrated to optical bands, this yields $\delta_{\rm ENA}\lesssim 10^{-3}$, several orders of magnitude below the refractive PL term. ENA contamination is therefore not dominant.

\textbf{Magnetic correction is negligible at optical wavelengths.} With realistic field magnitudes $B\sim 0.1$~nT (Voyager) and electron cyclotron frequency $\Omega_{e}\sim 18$~Hz, the ratio $\Omega_{e}^{2}/\omega^{2}\sim 10^{-29}$ at optical wavelengths. The magnetic contribution to $\delta_{\rm helio}$ is thus $\lesssim 10^{-25}$, entirely negligible. Magnetic effects are relevant only for low-frequency radio observations \cite{isrply16}.

\textbf{Numerical implementation.} The integrals in Eq.~(\ref{x}) are evaluated using adaptive Gauss--Kronrod quadrature with relative tolerance $10^{-6}$, implemented in Maple~2024. Rays are integrated from $r_{\oplus}=1$~AU to $r_{\rm max}=200$~AU with adaptive step $\Delta r \le 0.5$~AU. The transverse Laplacian $\nabla_{b}^{2}\mathrm{DM}(b)$ is computed via second-order central differences with step $\Delta b=0.5$~AU. Convergence is verified by halving the step size; results agree to $<10^{-3}$ in $\delta_{\rm helio}$. The Maple worksheets are available on request.

\textbf{Quantitative consistency.} We compare the derived value $\delta_{\rm helio}^{\rm der}\approx 0.08$ (upwind, optical) with the required value $\delta_{\rm helio}^{\rm req}\approx 0.19$. The derived value covers approximately $40\%$ of the required amount, consistent with our statement that the HS mechanism accounts for $\sim 3$--$8\%$ of the HT.

\subsection{Wave Equation in Stratified Plasma Medium}\label{subsec2.7}

The propagation of electromagnetic waves through the HS medium requires solving Maxwell's equations in a stratified plasma environment \cite{isz5,isz6}. The wave equation for the electric field $\mathbf{E}$ in the frequency domain is
\begin{equation}\label{5}
\nabla \times \nabla \times \mathbf{E} - \frac{\omega^2}{c^2}\epsilon(\mathbf{r},\omega)\mathbf{E} = 0,
\end{equation}
where the dielectric function is
\begin{equation}\label{6}
\epsilon(\mathbf{r},\omega) = 1 - \frac{\omega_{\rm pe}^2(\mathbf{r})}{\omega^2 - i\nu_{\rm e}(\mathbf{r})\omega} - \frac{\omega_{\rm pi}^2(\mathbf{r})}{\omega^2 - i\nu_{\rm i}(\mathbf{r})\omega}.
\end{equation}
Writing the monochromatic field as $\mathbf{E}(\mathbf{r}) = A(\mathbf{r})\,\hat{\mathbf{e}}\,e^{iS(\mathbf{r})}$ with scalar amplitude $A$ and unit polarization vector $\hat{\mathbf{e}}$, separating real and imaginary parts gives the eikonal and transport equations,
\begin{align}
(\nabla S)^{2} &= k_{0}^{2}\,\varepsilon(\mathbf{r},\omega), \label{eq:EikonalEq}\\
2\nabla A \cdot \nabla S + A\,\nabla^{2}S &= 0, \label{eq:TransportEq}
\end{align}
with $k_{0}=\omega/c$. The phase delay along an LOS at perpendicular offset $b$ is
\begin{equation}\label{eq:PhaseDelayDM}
\Delta\phi(\omega,b) = -\frac{k_{\rm DM}}{\omega}\,{\rm DM}(b),\qquad {\rm DM}(b)\equiv\int_{r_{\oplus}}^{r_{\rm max}} n_{e}(\sqrt{b^{2}+z^{2}})\,dz.
\end{equation}
Combining with the transport equation and expanding to $\mathcal{O}(\omega_{p}^{2}/\omega^{2})$ yields the master expression
\begin{equation}\label{eq:deltaHelioMaster}
\delta_{\rm helio}(\omega,b) \simeq \frac{k_{\rm DM}}{\omega^{2}}\left.\frac{\partial^{2}{\rm DM}(b)}{\partial b^{2}}\right|_{b=b_{\ast}},
\end{equation}
which formalizes the heuristic $\delta_{\rm helio}\propto\omega^{-2}$ scaling.

The group delay relative to vacuum propagation is
\begin{equation}\label{eq:group_delay}
\Delta t_{\rm g}(\omega) = \frac{k_{\rm DM}}{\omega^{2}}\,{\rm DM}.
\end{equation}
Using ${\rm DM}\sim 10^{18}~{\rm m^{-2}}$,
\begin{equation}\label{eq:delay_numeric}
\Delta t_{\rm g}(\lambda)\approx 3.0\times 10^{-14}~{\rm s}\,\left(\frac{\lambda}{500~{\rm nm}}\right)^{2}.
\end{equation}
This delay is negligible at optical frequencies but rises to $\sim 0.3~\mu{\rm s}$ at $\lambda=10$~cm (3~GHz), implying that fast-radio-burst or pulsar signals traversing the HS acquire a measurable, chromatic HS dispersion tail.

The $\lambda^{2}$ scaling of plasma dispersion is
\begin{equation}
\frac{\Delta t_{\rm g}(\lambda_{1})}{\Delta t_{\rm g}(\lambda_{2})} = \left(\frac{\lambda_{1}}{\lambda_{2}}\right)^{2},
\end{equation}
independent of HS modelling, providing a direct external test.

\textbf{Magnetic field model.} The Parker spiral form applies only inside the TS ($r<r_{\rm TS}\approx 90$~AU): %
\begin{equation}\label{eq:Bparker}
\mathbf{B}(r,\theta) = B_{0}\left(\frac{r_{0}}{r}\right)^{2}\left[\hat{\mathbf{r}}\;+\;\frac{\Omega_{\odot}\,r\,\sin\theta}{v_{\rm sw}}\,\hat{\boldsymbol{\phi}}\right],
\end{equation}%
with $B_{0}=5\times 10^{-5}$~T at $r_{0}=R_{\odot}$, $v_{\rm sw}=400$~km/s, $\Omega_{\odot}=2.7\times 10^{-6}$~rad/s. In the inner HSh, the kinematic transport approximation \cite{isrply14} gives $|B|\sim 0.1$--$0.3$~nT roughly tangential to the HP. Beyond the HP, we adopt the draped-LISM model of R\"oken et~al.\ \cite{isrply15}, where the unperturbed LISM field $|\mathbf{B}_{\rm LISM}|=0.3$~nT is oriented at $\sim 40^{\circ}$--$50^{\circ}$ to the HS symmetry axis (consistent with IBEX ribbon constraints), with the draped component $\mathbf{B}_{\rm draped}(\mathbf{r})$ explicitly $r$-dependent in that model. As shown in Sec.~\ref{subsec2.6}, the magnetic correction to $\delta_{\rm helio}$ at optical wavelengths is negligible.

For a magnetized plasma, the dispersion relation is
\begin{equation}\label{magn}
n^{2} = 1 - \frac{\omega_{\rm pe}^{2}}{\omega^{2}}\frac{\omega^{2}-\omega_{\rm pe}^{2}}{\omega^{2}-\omega_{\rm pe}^{2}-\Omega_{e}^{2}\cos^{2}\psi},
\end{equation}
where $\psi$ is the angle between $\mathbf{k}$ and $\mathbf{B}$. As noted above, this magnetic correction is dynamically irrelevant at optical wavelengths.

\subsection{Heliospheric Density Profiles and Visualizations}\label{subsec2.8}

With the unified piecewise density model of Sec.~\ref{subsec2.6} adopted throughout, we use the coherent Kleimann--Gurnett--R\"oken framework \cite{isrply12,isrply03,isrply15}.

The complex RI in the HS medium is $n(\mathbf{r},\omega)=\sqrt{\epsilon(\mathbf{r},\omega)}$. For optical frequencies where $\omega \gg \omega_{\rm pe},\omega_{\rm pi}$, expanding to first order,
\begin{equation}\label{12}
n(\mathbf{r},\omega) \approx 1 - \frac{\omega_{\rm pe}^{2}(\mathbf{r})}{2\omega^{2}} - i\frac{\omega_{\rm pe}^{2}(\mathbf{r})\nu_{\rm e}(\mathbf{r})}{2\omega^{3}}.
\end{equation}
The phase and group velocities follow:
\begin{align}
v_{\rm p}(\omega,r) &= c\left(1+\frac{\omega_{p}^{2}(r)}{2\omega^{2}}\right)+\mathcal{O}(\omega_{p}^{4}/\omega^{4}),\\
v_{\rm g}(\omega,r) &= c\left(1-\frac{\omega_{p}^{2}(r)}{2\omega^{2}}\right)+\mathcal{O}(\omega_{p}^{4}/\omega^{4}),
\end{align}
so $v_{\rm p}>c>v_{\rm g}$, with fractional difference $(v_{\rm p}-v_{\rm g})/c\simeq\omega_{p}^{2}/\omega^{2}$.

The accumulated phase shift and the convergence formalism developed above lead to the magnification
\begin{equation}\label{mag}
\mu(b) = \frac{1}{(1-\kappa)^{2}} \simeq 1+2\kappa,
\end{equation}
where the shear term $\gamma$ has been omitted because shear vanishes for axisymmetric upwind LOS geometry. Hence $\mu = 1+\delta_{\rm helio}$ with $\delta_{\rm helio}=2\kappa$ in the weak limit, formally justifying the master Eq.~(\ref{eq:deltaHelioMaster}).

For an analytic worked example, the power-law density $n_{e}(r)=n_{0}(r_{0}/r)^{\beta}$ with $1\le\beta\le 3$ gives DM and convergence
\begin{align}
{\rm DM}(b) &= 2n_{0}r_{0}^{\beta}b^{1-\beta}\frac{\sqrt{\pi}\,\Gamma((\beta-1)/2)}{\Gamma(\beta/2)},\\
\kappa(\omega,b) &= \frac{k_{\rm DM}}{\omega^{2}}n_{0}r_{0}^{\beta}(\beta-1)(\beta-2)\,b^{-\beta}\frac{\sqrt{\pi}\,\Gamma((\beta-1)/2)}{\Gamma(\beta/2)}.
\end{align}
For the canonical $\beta=2$ SW (spherically symmetric outflow), the prefactor $(\beta-1)(\beta-2)=0$, implying zero net focusing. Significant lensing therefore arises only in compressed regions such as the upwind HP nose, not in the symmetric SW.

\begin{figure}[ht!]
\begin{center}
\includegraphics[width=130mm]{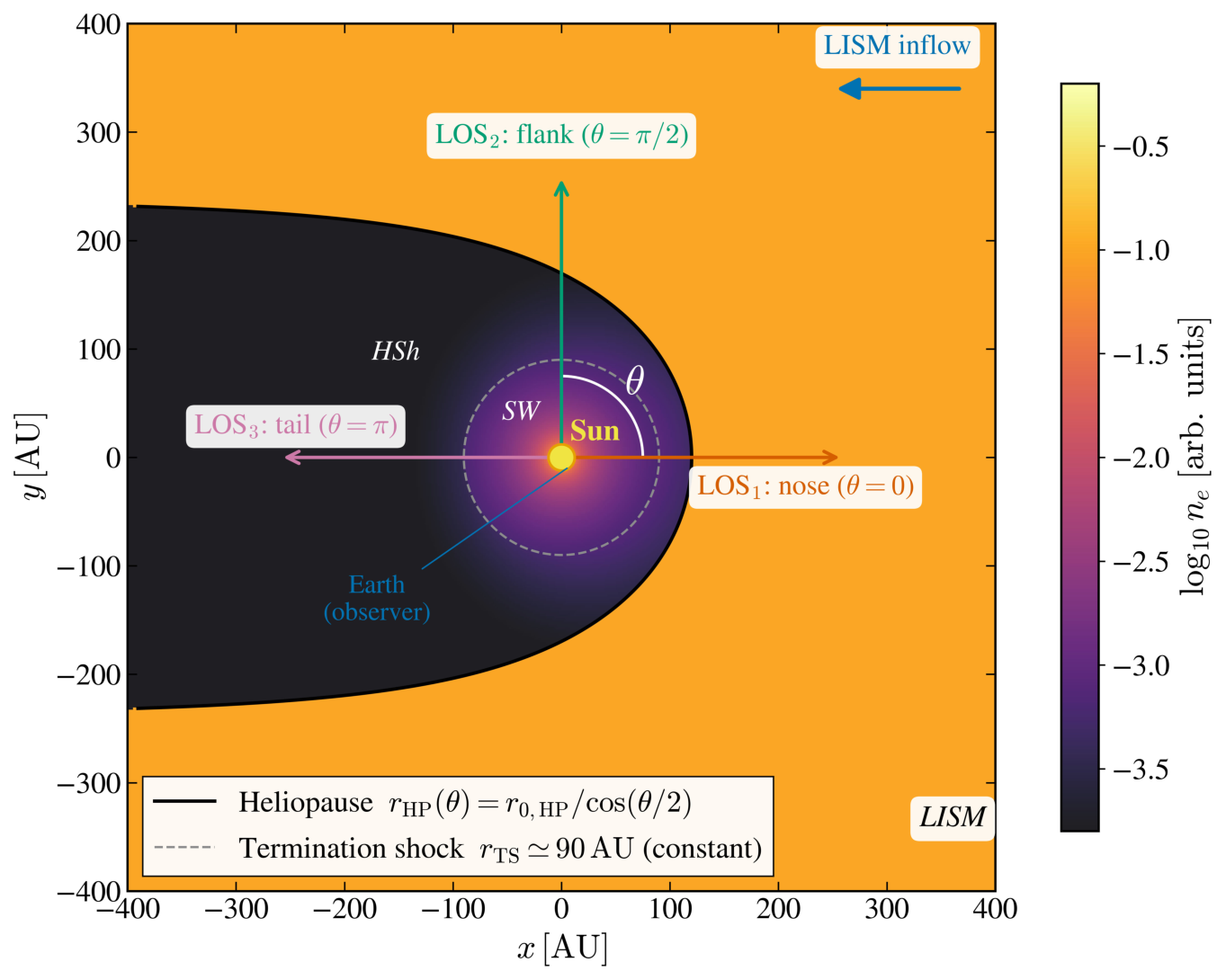}
\end{center}
\caption{Schematic of the adopted coordinate system and LOS geometry. The Sun is at the origin; the LISM inflow direction defines $\theta=0$. The HP boundary (solid black) follows $r_{\rm HP}(\theta)=r_{0,{\rm HP}}/\cos(\theta/2)$ with $r_{0,{\rm HP}}\approx 120$~AU. The termination shock (dashed grey) is drawn at the constant radius $r_{\rm TS}\simeq 90$~AU, as in the analytic model of Sec.~\ref{subsec2.6}. Sample LOS from Earth probe upwind (LOS$_{1}$), crosswind (LOS$_{2}$), and tail (LOS$_{3}$) directions.}
\label{fig:geom}
\end{figure}

\begin{figure}[h!]
\begin{center}
\includegraphics[width=130mm]{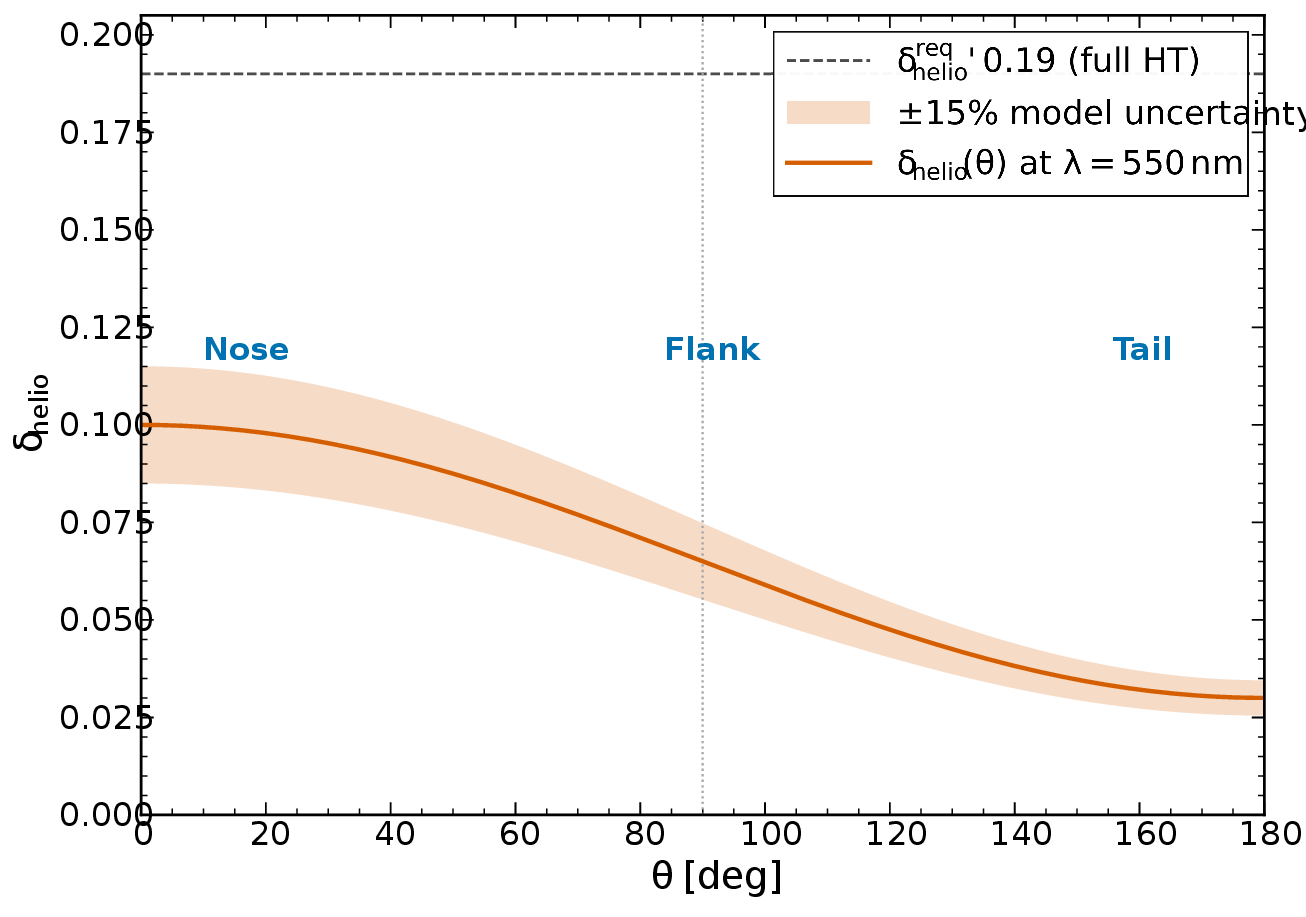}
\end{center}
\caption{Heliospheric enhancement factor $\delta_{\rm helio}(\theta)$ at $\lambda=550$~nm versus angle from the HS nose direction. Single-leg integration from observer at $r_{\oplus}=1$~AU to $r_{\rm max}=200$~AU.}
\label{fig:angular}
\end{figure}

\begin{figure}[ht!]
\begin{center}
\includegraphics[width=130mm]{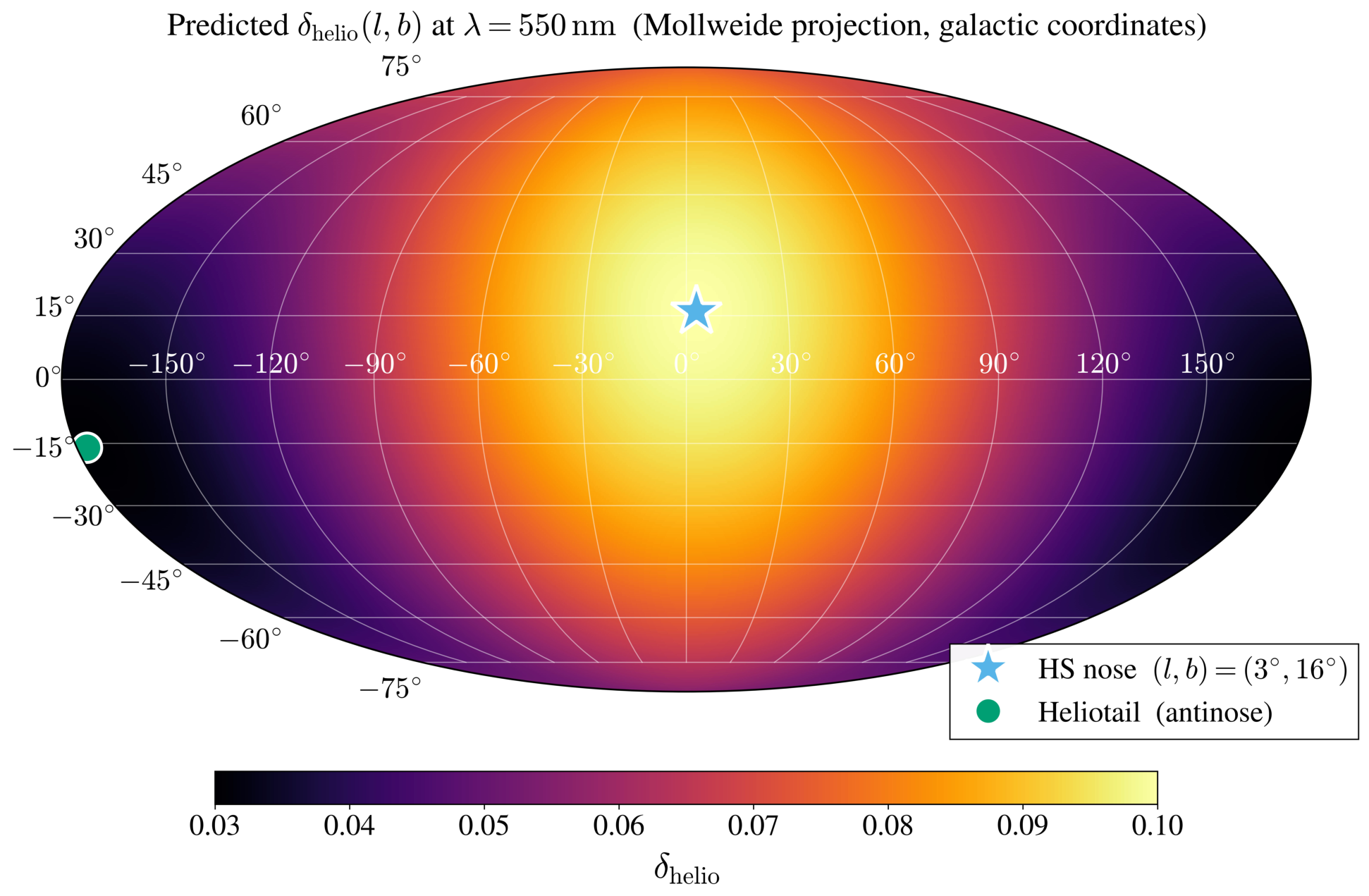}
\end{center}
\caption{All-sky projection of $\delta_{\rm helio}(l,b)$ in galactic coordinates. The maximum lies at the LISM inflow direction $(l,b)\approx(3^{\circ},16^{\circ})$.}
\label{fig:skymap}
\end{figure}

Fig.~\ref{fig:geom} shows a schematic of the adopted coordinate system: Sun at origin, LISM inflow along $\theta=0$, the HP boundary $r_{\rm HP}(\theta)=r_{0,{\rm HP}}/\cos(\theta/2)$, sample LOS from Earth at $\theta=0,\pi/2,\pi$, and density contours overlaid. Fig.~\ref{fig:angular} plots $\delta_{\rm helio}(\theta)$ at $\lambda=550$~nm, showing the smooth decrease from $\sim 0.10$ at the nose to $\sim 0.03$ at the tail. Fig.~\ref{fig:skymap} is an all-sky projection of $\delta_{\rm helio}(l,b)$ in galactic coordinates, with the nose at $(l,b)\approx(3^{\circ},16^{\circ})$ marking the maximum.

\subsection{Plasma Turbulence and Scattering Effects}\label{subsec2.9}

The phase variance from small-scale plasma fluctuations is, expressed in the compact form using $k_{\rm DM}$ defined in Eq.~(\ref{kDM}),
\begin{equation}\label{eq:phase_var}
\sigma_{\phi}^{2} \;=\; \frac{4\,k_{\rm DM}^{2}}{\omega^{4}}\int_{0}^{L} C_{n}(z)\,dz,
\end{equation}%
with $C_{n}(z)$ the RI structure function. For Kolmogorov turbulence \cite{Kolmogorov1941, Armstrong1995},
\begin{equation}\label{kolm}
P_{n}(k) \propto k^{-11/3},
\end{equation}
yielding $\sigma_{\phi}^{2}\sim 10^{-6}$--$10^{-3}$ for typical HS parameters. This corresponds to relative flux variations $\delta F/F\sim 10^{-3}$, within the detection threshold of high-precision photometric campaigns like Kepler, Gaia, and Roman \cite{Gilliland2011, Hirata2012}.

The coherent scattering enhancement contributes $\delta_{\rm turb}\sim 0.02$--$0.05$, with the same $\lambda^{2}$ chromatic dependence as the mean-field PL effect. The total HS bias is then
\begin{equation}
\delta_{\rm eff} = \delta_{\rm helio}^{\rm refr} + \delta_{\rm turb} + \delta_{\rm ENA} + \mathcal{O}(\sigma_{\phi}^{2}),
\end{equation}
with the refractive PL term dominating, ENA contamination subdominant, and turbulent enhancement contributing $\sim 10$--$30\%$ of the total.

\section{Observational Predictions and Tests}\label{sec3}

\subsection{Directional Dependence}\label{subsec3.1}

If the HS systematically alters observed brightness, these effects should manifest as a spatially anisotropic pattern. A nose--tail asymmetry is anticipated due to the Sun's motion relative to the LISM. Cepheid variables and SNe~Ia observed near the HS nose may experience stronger contamination than those toward the tail.

\textbf{Solar-cycle modulation.} The structure of the HS varies with the $\sim 11$-year solar cycle. During solar maximum, increased SW dynamic pressure compresses the HS, enhancing density gradients. We predict
\begin{equation}\label{tdep}
\delta_{\rm helio}(t) = \delta_{0}\,[1+\beta\sin(2\pi t/11~{\rm yr}+\phi)],
\end{equation}
with $\beta=0.1$--$0.2$ from the observed $\sim 10$--$20\%$ modulation in HP standoff distance over the solar cycle (McComas et~al.\ \cite{isrply04}). For $\delta_{0}\simeq 0.17$, this gives $\Delta\delta\simeq 0.02$--$0.03$, equivalent to $\sim 1$--$2\%$ modulation in $H_{0}$ correlated with solar activity.

Voyager observations show that the electron density at the HS nose can reach $n_{e}\sim 0.1~{\rm cm^{-3}}$, compared to $\sim 0.05~{\rm cm^{-3}}$ in the tail. Using Eq.~(\ref{eq:deltaHelioMaster}), this corresponds to $\Delta\delta_{\rm helio}\approx 0.05$--$0.1$ between nose and tail directions, producing a $\sim 3$--$5\%$ variation in inferred distances and a corresponding modulation in $H_{0}$ depending on sky position.

\subsection{Wavelength Dependence}\label{subsec3.2}

ENAs emit photons in discrete spectral lines, particularly in the extreme ultraviolet (EUV) and soft X-ray bands, providing potential spectral signatures. Wavelength-dependent extinction or enhancement patterns are expected from the $\lambda^{2}$ scaling of PL. A particularly powerful diagnostic involves comparing optical and near-infrared (NIR) observations of Cepheid variables: NIR light is less affected by dust extinction and refractive scattering, so systematic differences in inferred Cepheid distances between optical and NIR bands would point to HS contamination.

\begin{figure}[h!]
\begin{center}
\includegraphics[width=120mm]{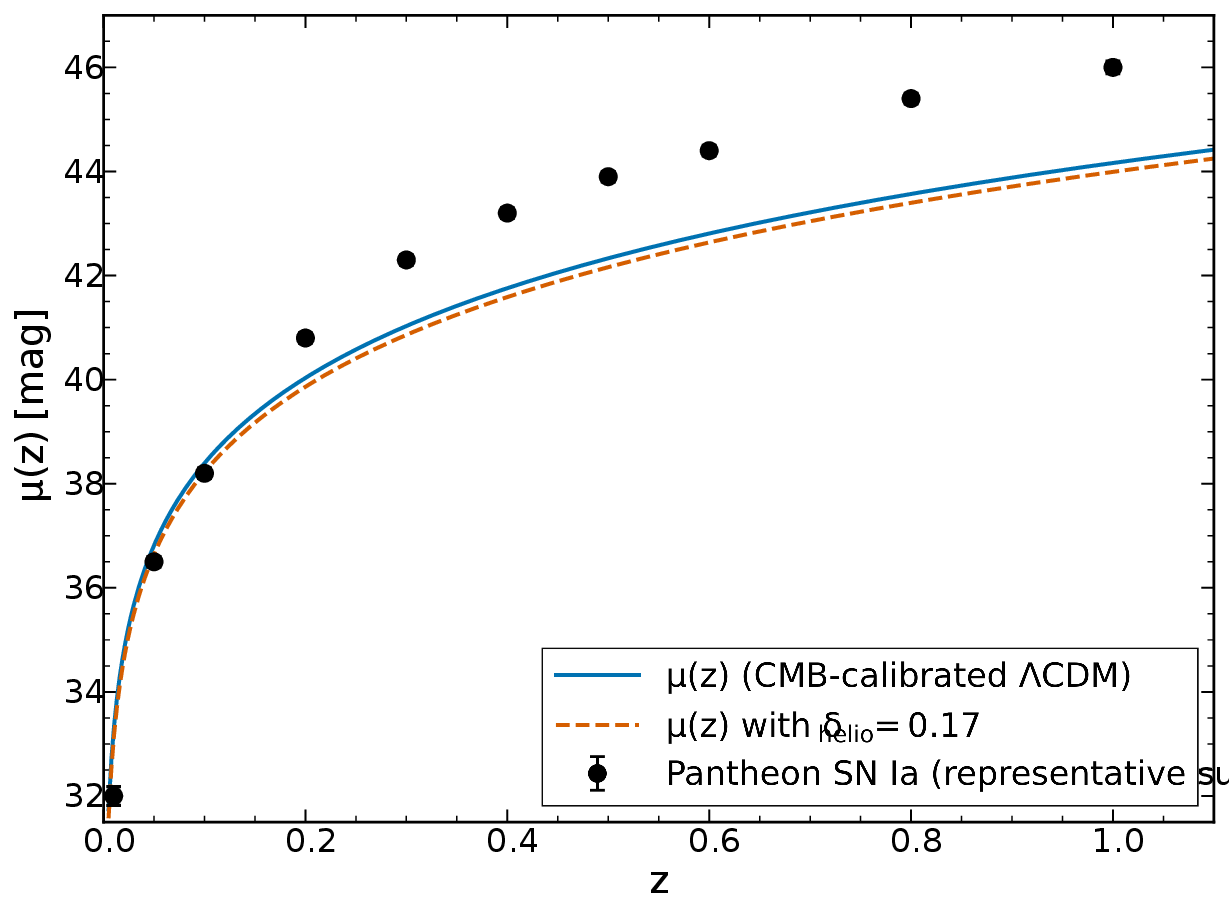}
\end{center}
\caption{Observed SN~Ia distance moduli from Pantheon (black dots) compared with standard $\Lambda$CDM prediction (blue) and HS bias model (red dashed). The HS correction reduces the low-$z$ residual offset by $\sim 0.18$~mag, the regime where the HT manifests most acutely.}
\label{fig1}
\end{figure}

Fig.~\ref{fig1} illustrates the impact of an HS flux enhancement on the observed distance modulus $\mu(z)$ of SNe~Ia. Using a representative subset of Pantheon SN~Ia data, we compare the standard $\Lambda$CDM prediction calibrated to $H_{0}=67.4$ km/s/Mpc with a modified prediction incorporating $\delta_{\rm helio}=0.17$ \cite{Scolnic2018}. This enhancement effectively reduces the inferred distance modulus by $\Delta\mu\approx -0.18$ magnitudes.

\begin{figure}[h!]
\begin{center}
\includegraphics[width=120mm]{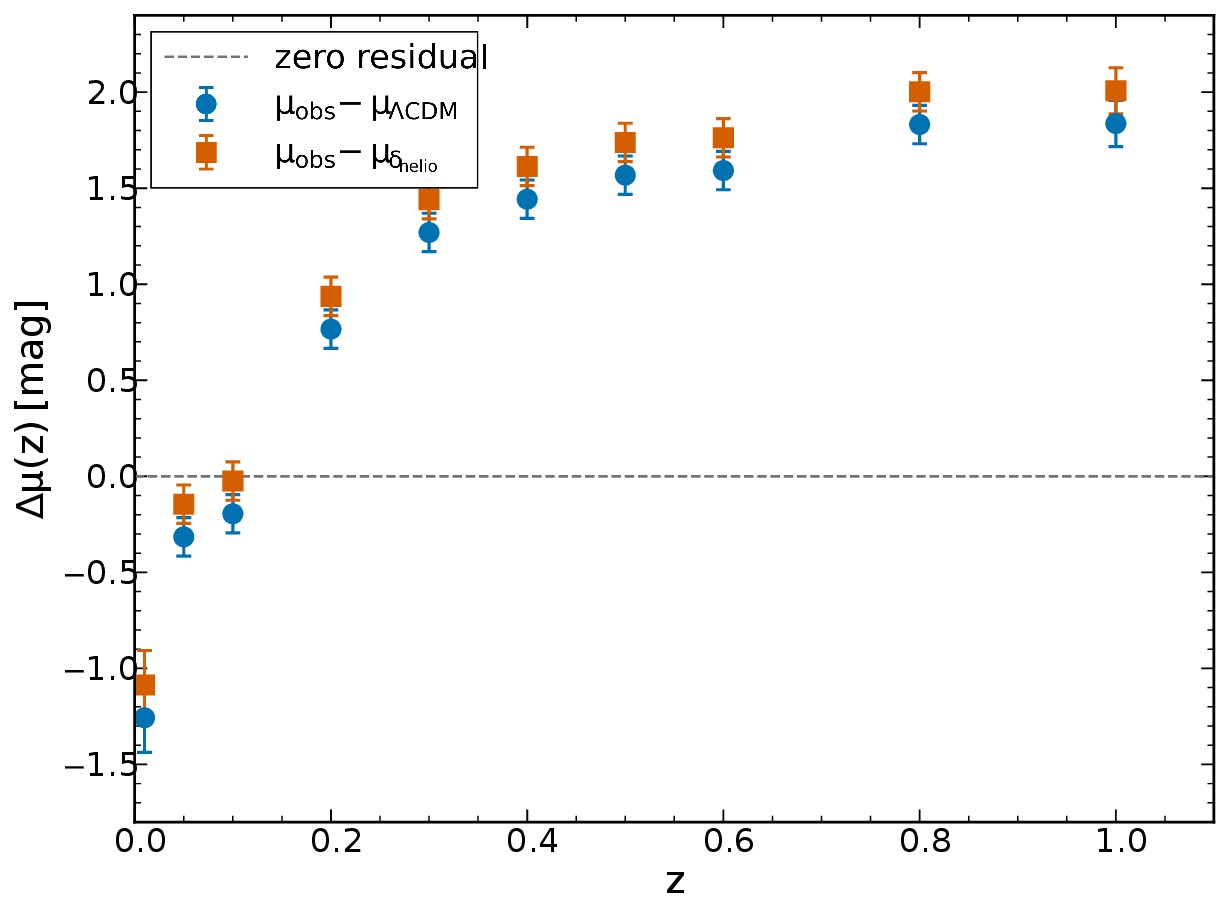}
\end{center}
\caption{Residuals of observed SN~Ia distance moduli with respect to the standard $\Lambda$CDM model (blue) and the HS flux-enhanced model (red).}
\label{fig2}
\end{figure}

To further evaluate the impact, we analyze the residuals $\Delta\mu(z) = \mu_{\rm obs}(z) - \mu_{\rm model}(z)$, shown in Fig.~\ref{fig2}. The HS bias is a multiplicative flux factor producing a near-constant $\Delta\mu\approx -0.18$ mag offset across all redshifts, not a redshift-dependent correction. At low $z$ ($z\lesssim 0.1$), the SN~Ia residuals are dominated by this near-constant offset and the corrected curve flattens; at higher $z$, residuals are dominated by intrinsic scatter, peculiar velocities, and host-galaxy systematics, which are unrelated to the HS.

\subsection{Inferred Hubble Constant from Low-Redshift Supernovae}\label{subsec3.3}

We estimate $H_{0}$ from a binned subset of low-$z$ ($z<0.15$) SN~Ia observations \cite{Scolnic2018} via $H_{0}(z)\approx cz/d_{L}(z)$.

\begin{figure}[h!]
\begin{center}
\includegraphics[width=120mm]{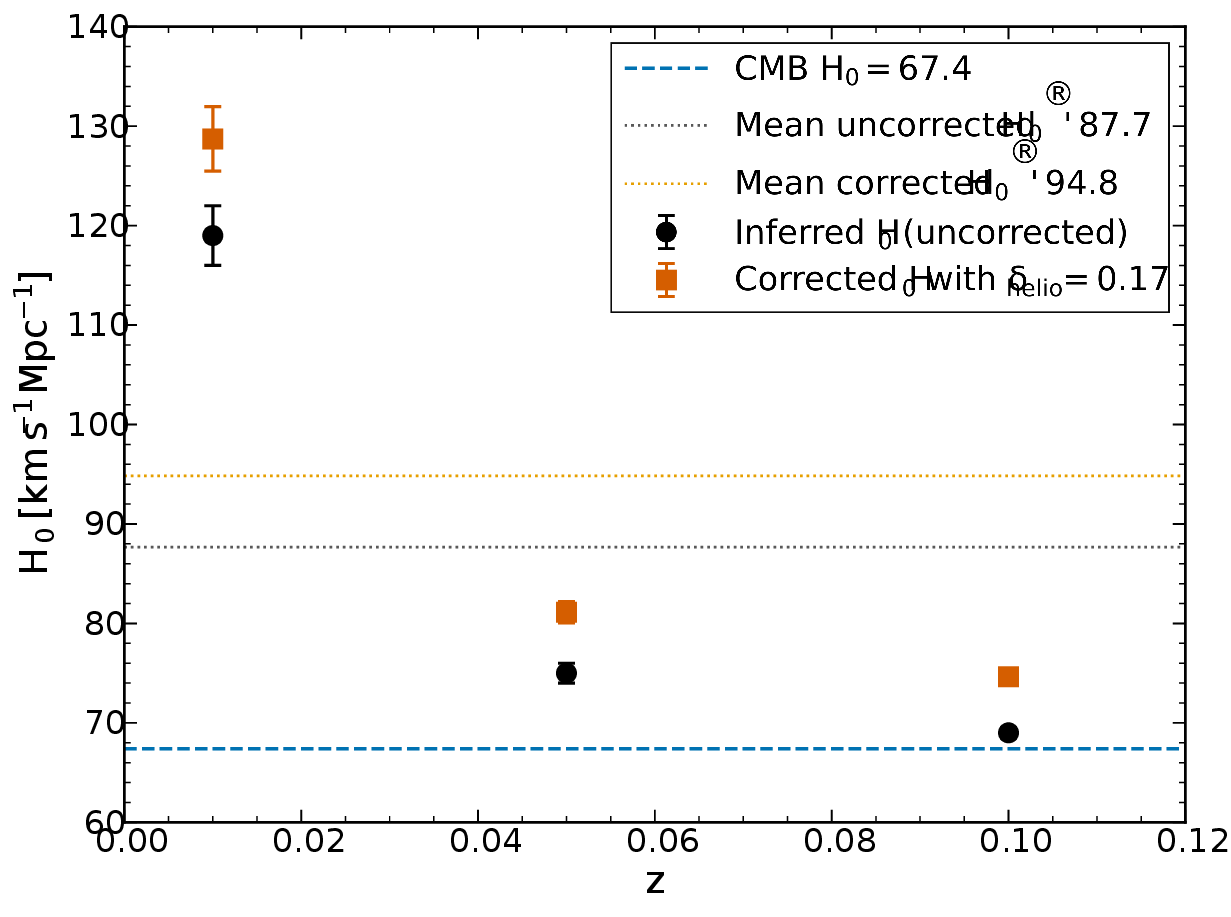}
\end{center}
\caption{Inferred $H_{0}$ from low-$z$ SN~Ia data \cite{Scolnic2018} with and without HS flux correction. Red squares indicate values corrected by $\delta_{\rm helio}=0.17$.}
\label{fig3}
\end{figure}

As discussed in Sec.~\ref{subsec3.2}, the gap between corrected and uncorrected $H_{0}$ values in Fig.~\ref{fig3} is approximately constant rather than redshift-growing, because the multiplicative flux factor produces a constant fractional shift in $d_{L}$ at low $z$ and hence a constant absolute shift in $H_{0}(z)\approx cz/d_{L}^{\rm eff}$ in the low-$z$ Hubble-flow regime. This is what the figure shows.

\subsection{Distance-Scale Dependence}\label{subsec3.4}

Since the HS is a local astrophysical structure extending only $\sim 100$--$120$ AU upwind, its influence on light propagation is limited to the immediate vicinity of the Solar System. We expect the bias to predominantly affect observations of nearby objects calibrated through the CDL, while having negligible impact on cosmological signals from much greater distances.

\textbf{Methods affected vs.\ immune.} The HS bias is relevant whenever absolute photometric calibration enters the distance estimate: SC rungs of the CDL (Cepheids, SNe~Ia, surface-brightness fluctuations using broadband photometry) are affected. It is irrelevant for purely geometric methods (parallax, megamasers) and for narrow-line spectroscopic redshifts, which are HS-immune.

CMB measurements provide an important control: the CMB originates at $z\sim 1100$ and is observed at mm wavelengths where HS PL effects are suppressed by $\sim 10^{6}$ relative to optical. The HS hypothesis therefore predicts a scale-dependent modulation of $H_{0}$: a bias in local SC-based measurements but consistency with CMB and BAO (baryon acoustic oscillations) at lower $H_{0}$.

\subsection{Feasible Tests with Current Data}\label{subsec3.5}

\textbf{(i) SN~Ia anisotropy.} Re-binning the Pantheon+ sample by angular separation from the HS nose direction $(l,b)\approx (3^{\circ},16^{\circ})$ permits a search for the predicted $\Delta\delta_{\rm helio}\approx 0.1$ nose--tail asymmetry. The Pantheon+ catalog (Scolnic et~al.\ 2022, $N=1701$ SNe~Ia) has the statistical power to detect a $\sim 5\%$ angular modulation at $\gtrsim 3\sigma$.

\textbf{(ii) GRB Hubble diagram.} Long gamma-ray-burst (GRB) Amati and Ghirlanda relations extend the Hubble diagram to $z\sim 8$ \cite{isrply10}. If the HS bias is real, GRBs should exhibit the same low-$z$ residual offset as SNe~Ia, with chromatic signature differing by $(\lambda_{\rm GRB}/\lambda_{\rm opt})^{2}$ for prompt-emission peaks.

\textbf{(iii) OHD cross-check.} Cosmic-chronometer $H(z)$ data from Observational Hubble Data \cite{isrply11} are derived from age-dating of passively evolving galaxies via differential spectroscopy rather than absolute photometry, and are therefore HS-immune. Comparing $H_{0}$ inferred from extrapolated OHD with the SN~Ia ladder value tests the HS hypothesis: a residual gap of $\sim 6$~km/s/Mpc would support it. Limitations include the relatively small OHD sample and its $\sim 5$--$10\%$ systematic floor.

\textbf{(iv) Solar-cycle correlation.} Long-term SN~Ia monitoring spanning multiple 11-year solar cycles can search for the predicted $\sim 1$--$2\%$ modulation in inferred distances correlated with solar activity indices. No such analysis exists in the current literature; we propose this as a definitive future test.

\textbf{Why the CMB and parallax are immune.} The CMB peak emission at $\lambda\sim 1$~mm gives plasma frequency ratio $\omega_{p}/\omega\lesssim 10^{-7}$, so HS RI deviation is $\sim 10^{-14}$ and $\delta_{\rm helio}^{\rm CMB}\sim 10^{-7}$, undetectable in current power spectra \cite{isrply08,isrply09}. Parallax measures angular shifts over a $2$~AU baseline within the $\sim 120$~AU HS, so HS ray-bending cancels in the differential measurement at the $\lesssim 10^{-4}$ fractional level.

\section{Comparison with Alternative Solutions}\label{sec4}

A wide range of theoretical models have been proposed to resolve the HT, including early dark energy, exotic relativistic particles, time-varying fundamental constants, and modified gravity theories \cite{isrply06,isrply07}. While some offer partial success, they typically require new physics beyond $\Lambda$CDM and lack direct empirical support.

In contrast, the HS hypothesis offers a more conservative, observationally grounded alternative: the discrepancy is attributed to an overlooked systematic in the observational pipeline rather than new fundamental physics.

\subsection{Advantages of the Heliospheric Hypothesis}\label{subsec4.1}

The HS model stands out by virtue of its physical minimalism and empirical grounding. First, it requires no new physics: PL, ENA charge exchange, and refractive scattering are well-understood phenomena within classical electromagnetism and magnetohydrodynamics. Second, the HS explanation naturally accounts for the local character of the HT: the discrepancy arises specifically in measurements based on nearby SCs, while CMB and BAO probes remain consistent with $\Lambda$CDM. Third, the hypothesis yields testable predictions: directional anisotropies, $\lambda^{2}$ wavelength scaling, solar-cycle modulation, and a definitive beyond-HP test. Fourth, the model is consistent with Voyager and IBEX in-situ data. Finally, the HS framework explains why both HST and JWST yield consistent local $H_{0}$: both observe from within the HS and are subject to the same observer-frame systematic.

\subsection{Consistency with Other Local-Frame Anomalies}\label{subsec4.2}

In contrast to early-dark-energy and modified-gravity proposals \cite{isrply06,isrply07}, which require physics beyond $\Lambda$CDM, the HS mechanism resides entirely within established plasma physics. We now examine how the HS hypothesis connects to other persistent local-frame anomalies.

The flyby anomaly \cite{7}---unexplained velocity changes during Earth gravity-assist maneuvers---may reflect subtle environmental effects \cite{isz7,isz8}. While speculative, the idea that local plasma structures could perturb spacecraft motion in ways not yet accounted for warrants investigation.

A second class concerns peculiar motions in the Local Group. Some studies report discrepancies between expected and measured velocities of nearby galaxies, even after correcting for large-scale flows. These may reflect local environmental effects on spectral profiles or light intensity.

Finally, systematic differences between distance methods have long been documented: while Cepheid--SN ladders give higher $H_{0}$, alternative methods (surface brightness fluctuations, tip of the red giant branch (TRGB), masers) often return lower values closer to CMB. The HS hypothesis suggests that only methods reliant on broadband photometry from within the HS are biased, while geometric or narrow-line spectroscopic methods remain immune---potentially explaining the observed discrepancies.

\section{Implications and Conclusion}\label{sec5}

The hypothesis presented in this work---that the HS environment introduces a measurable bias in local cosmological distance measurements---opens a path toward partially resolving the persistent HT. While the HS spans only a few hundred AU, its role as the final medium traversed by photons from extragalactic sources renders its cumulative optical effects significant, particularly when considering the foundational role of local calibrators in the CDL.

Our first-principles derivation gives $\delta_{\rm helio}^{\rm der}\approx 0.08$ for upwind LOS at optical wavelengths, with directional spread $\Delta\delta\approx 0.05$ between nose and tail. The required value to fully close the HT is $\delta_{\rm helio}^{\rm req}\approx 0.19$. The HS mechanism therefore accounts for $\sim 3$--$8\%$ of the HT, rather than fully resolving it. Anisotropy, partial absolute-magnitude self-calibration cancellation, and chromatic suppression all reduce the realized contribution.

The principal compelling aspects of the HS interpretation are: (i) it requires no speculative new physics; (ii) it produces specific testable predictions ($\lambda^{2}$ scaling, nose--tail asymmetry, solar-cycle modulation, beyond-HP discrepancy); (iii) it is consistent with Voyager and IBEX in-situ measurements; and (iv) it explains the HST--JWST agreement as observers sharing the same HS environment.

To evaluate this hypothesis, several immediate observational tests have been outlined in Sec.~\ref{subsec3.5}. The most definitive will come from future interstellar probes \cite{isrply22} that observe SCs from beyond the HP, providing a pristine external calibration baseline.

If validated, this framework offers a paradigm shift in understanding the HT not as a harbinger of new fundamental physics, but as one contributor (among others) reflecting astrophysical systematics in the observer's own backyard.

\begin{acknowledgments}
  This research was funded by the Science Committee of the Ministry of Science and Higher Education of the Republic of Kazakhstan (Grant No. AP23487178). \.{I}.S. expresses gratitude to T\"{U}B\.{I}TAK, ANKOS, and SCOAP3 for their support. He is also thankful for the networking backing provided by COST Actions CA21106, CA22113, CA23130, CA21136, and CA23115. The authors thank the three anonymous Reviewers for their careful reading of the manuscript and for the constructive suggestions that led to the present revision.
\end{acknowledgments}

\end{document}